\documentclass[referee]{aa} 
\usepackage{graphicx}
\def\lesssim{\mathrel{\hbox{\rlap{\hbox{\lower4pt\hbox{$\sim$}}}\hbox{$<$}}}}
\def\gtrsim{o.fits.\mathrel{\hbox{\rlap{\hbox{\lower4pt\hbox{$\sim$}}}\hbox{$>$}}}}
\newcommand{\mincir}{\raise
-2.truept\hbox{\rlap{\hbox{$\sim$}}\raise5.truept
\hbox{$<$}\ }}
\newcommand{\magcir}{\raise
-2.truept\hbox{\rlap{\hbox{$\sim$}}\raise5.truept
\hbox{$>$}\ }}
\newcommand{\siml}{\raise -2.truept\hbox{\rlap{\hbox{$\sim$}}\raise5.truept
\hbox{$<$}\ }}
\newcommand{\simg}{\raise -2.truept\hbox{\rlap{\hbox{$\sim$}}\raise5.truept
\hbox{$>$}\ }}
\newcommand{\be}{\begin{equation}}
\newcommand{\ee}{\end{equation}}
\newcommand{\ba}{\begin{eqnarray}}
\newcommand{\ea}{\end{eqnarray}}

\newcommand {\ks} {km~s$^{-1} \;$}

\begin{document}
   \title{Optical Luminosity Functions of the Abell Galaxy Cluster ABCG\,209 at z=0.21 
\thanks{Based on observations collected at the European Southern
Observatory, Chile (Proposal ESO NM-0 68.A-0116)}}

\author{A. Mercurio\inst{1}
\and M. Massarotti \inst{2} 
\and P. Merluzzi \inst{2} 
\and M. Girardi \inst{1} 
\and F. La Barbera \inst{2}
\and G. Busarello\inst{2} 
}                     
\offprints{A. Mercurio}          
\institute{Dipartimento di Astronomia, Universit\`{a} degli Studi di Trieste,
Via Tiepolo 11, I-34100 Trieste, Italy\\
\email{mercurio,girardi@ts.astro.it}
\and   INAF - Osservatorio Astronomico di Capodimonte,
via Moiariello 16, I-80131 Napoli, Italy\\
}
\date{Received: date 27-02-03/ Revised version: date 12-05-03 }
%

\abstract
{We derive the luminosity functions in three bands (BVR) for
the rich galaxy cluster ABCG\,209 at z=0.21. The data cover an area of
$\sim$ 78 arcmin$^2$ in the B and R bands, while a mosaic of three
pointings was obtained in the V band, covering an area of
approximately 160 arcmin$^2$. The galaxy sample is complete to B =
22.8 ($\mathrm{N_{gal}}$ = 339), V = 22.5 ($\mathrm{N_{gal}}$ = 1078)
and R = 22.0 ($\mathrm{N_{gal}}$ = 679). Although the fit of a
single Schechter function cannot be rejected in any band, the
luminosity functions are better described by a sum of two Schechter
functions for bright and faint galaxies, respectively. There is an
indication for a presence of a dip in the luminosity functions in the
range V = 20.5--21.5 and R = 20.0--21.0.  We find a marked luminosity
segregation, in the sense that the number ratio of bright--to--faint
galaxies decreases by a factor 4 from the center to outer regions. Our
analysis supports the idea that ABCG\,209 is an evolved cluster,
resulting from the merger of two or more sub--clusters.

} 
%
\titlerunning{Optical Luminosity Functions of ABCG\,209 at z$\sim$0.21}

\maketitle
\section{Introduction}
\label{intro}

Galaxy luminosity function (LF) is a powerful tool to constrain galaxy
formation and evolution, since it is directly related to the galaxy
mass function and hence to the spectrum of initial perturbations.
Hierarchical clustering models predict a mass distribution
characterised by a cut--off above a given mass M$^*$ and well
described by a power law at low masses (Press \& Schechter
\cite{pre74}). Starting from these results Schechter (\cite{sch76})
analysed the luminosity distribution of 14 galaxy clusters observed by
Oelmer (\cite{oel74}), by introducing an analytical description in the
form:

\begin{equation}
\mathrm{
\phi(L)dL =
\phi^*\left(\frac{L}{L^*}\right)^{\alpha}
e^{-\left(\frac{L}{L^*}\right)} d\left(\frac{L}{L^*}\right) \ .
}
\label{eqsch}
\end{equation}

\noindent
In the Schechter function $\phi^*$ is a normalization density and the
shape of the LF is described by L$^*$, a characteristic cut--off
luminosity, and $\alpha$, the faint--end slope of the distribution. He
used this function to describe the global LF of all galaxy types and
suggested the value $\alpha = -5/4$.

Although investigated in several works, the universality of the LF
faint--end slope is still controversial. The value of the faint--end
slope turns out to be $\alpha\sim-1$ for field galaxies (e.g.,
Efstathiou et al.~\cite{efs88}; Loveday et al.~\cite{lov92}), while
clusters and groups seem to have steeper slopes, -1.8 $< \alpha<$ -1.3
(e.g., De Propris et al. \cite{dep95}; Lumsden et al. \cite{lum97};
Valotto et al. \cite{val97}), suggesting the presence of a larger
number of dwarf galaxies (but see also e.g., Lugger \cite{lug86};
Colless \cite{col89}; Trentham \cite{tre98a}). Changes in the slope of
the faint--end of the LF in clusters can be related to environmental
effects. An increase of the steepening of the LF faint--end in the
cluster outer regions was actually observed (Andreon \cite{and01}) and
explained taking into account that the various dynamical processes
which can destroy dwarf galaxies act preferentially in the
higher--density cores.

Lopez--Cruz et al. (\cite{lop97}) showed that clusters with a flat LF
($\alpha\sim-1$) are a homogeneous class of rich clusters with a
single dominant galaxy, symmetrical single peaked X--ray emission and
high gas masses. Irregular clusters have a steeper faint--end, in
particular, the LFs of ABCG\,1569 and Coma which present
substructures, can be suitably fitted with the sum of two Schechter
functions with $\alpha=-1$ and $\alpha \geq -1.4$ (Lopez--Cruz et al.
\cite{lop97}). Trentham (\cite{tre97}) also suggested that the
faint--end slope of the LF flattens as clusters evolve because of the
destruction of dwarf galaxies by merging with giants galaxies.

The density of the environment seems to affect the distribution of
galaxy luminosity in the sense that Schechter fits are poor for
galaxies in dense environment, where there is an indication of a dip
(e.g., Driver et al. \cite{dri94}; Biviano et al. \cite{biv95}; Wilson
et al. \cite{wil97}; Molinari et al. \cite{mol98}; Garilli et
al. \cite{gar99}; N$\mathrm{\ddot{a}}$slund et al. \cite{nas00}; Yagi
et al. \cite{yag02}). Trentham \& Hodgking (\cite{tre02}) identified
two types of galaxy LF, one for dynamically evolved regions
(i.e. region with a high elliptical galaxy fraction, a high galaxy
density, and a short crossing time), such as Virgo cluster and Coma
cluster, and one for unevolved regions, such as the Ursa Major cluster
and the Local Group. A dip is present in the LF of Virgo and Coma and
is absent in LFs of Ursa Major and Local Group.

The differences in shape of the LFs from cluster to cluster could be
explained assuming that the total LF is the sum of type specific
luminosity functions (hereafter TSLFs), each with its universal shape
for a specific type of galaxies (Binggeli et al. \cite{bin88}). The
total LF then assumes a final shape which can be different from
cluster to cluster according to the mixture of different galaxy
types. Therefore, different mixtures of galaxies, induced by
cluster--related processes, may be at the origin of the presence and
of the different shape of dips seen in cluster LFs and may be
responsible for the differences seen in the total LFs among field,
groups and galaxy clusters.

Indeed, dips are found in several clusters, occurring roughly at
the same absolute magnitude (M$\mathrm{_{R,dip}} \sim$ -19.4 or $\sim$
M$^*$ + 2.5), within a range of about one magnitude, suggesting that
clusters have comparable galaxy population. However the dips may have
different shapes, and also depend on the cluster region. This could be
related to the relative abundances of galaxy types, which depend on
the global properties of each cluster and on the local density (Durret
et al. \cite{dur99}).

In this work we study the Abell galaxy cluster ABCG\,209 at z = 0.21
(Kristian et al. \cite{kri78}; Wilkinson \& Oke \cite{wil78}; Fetisova
\cite{fet81}; Mercurio et al. \cite{mer03}, hereafter Paper I) which
is a rich, X--ray luminous, and massive cluster (richness class R = 3,
Abell et al. \cite{abe89}; $\mathrm{L_X(0.1-2.4\ keV)\sim14\cdot
10^{44} \ h_{50}^{-2}\ erg\ s^{-1}}$, Ebeling et al. \cite{ebe96};
$\mathrm{T_X\sim10\ keV}$, Rizza et al. \cite{riz98}; M($<$
R$\mathrm{_{vir}}$) = 1.6--2.2 10$^{15}$ $\mathrm{h^{-1}_{100} \ M_{\odot}}$,
Paper I).

The cluster shows an elongation and asymmetry in the X--ray emission
with two main clumps (Rizza et al. \cite{riz98}), but no strong
cooling flow is detected. The dynamical analysis presented in Paper I
showed that ABCG\,209 is characterized by a very high value of
the line of sight velocity dispersion:
$\mathrm{\sigma_v=1250}$--$1400$ km s$^{-1}$ and by a preferential
SE--NW direction as indicated by: a) the presence of a velocity
gradient in the velocity field; b) the elongation in the spatial
distribution of colour--selected cluster members; c) the elongation of
the X--ray contour levels in the Chandra image; d) the elongation of
the cD galaxy. There is significant evidence of substructure, as shown
by the Dressler \& Schectman test. The two--dimensional distribution
of the colour--selected members shows a strong luminosity
segregation.  Furthermore, the young dynamical state is also
indicated by the possible presence of a radio halo (Giovannini et
al. \cite{gio99}), possibly a remnant of a recent cluster merger
(Feretti et al. \cite{fer00}).

This observational scenario suggests that ABCG\,209 is presently
undergoing strong dynamical evolution with the merging of two or more
sub--clumps along the SE--NW direction, but could not allow us to
discriminate between two alternative pictures (Paper I). The merging
might be either in a very early dynamical state, where the clumps are
still in a pre--merging phase, or in a more advanced state, where
luminous galaxies trace the remnant of the core--halo structure of a
pre--merging clump hosting the cD galaxy.

In order to further investigate the cluster dynamical state and to
discriminate between the previous pictures, we derived the LFs by
using new photometric data for ABCG\,209 based on ESO--NTT imaging in
the B, V and R wavebands. The new photometric data are presented in
Sect. \ref{sec:2}. In Sect. \ref{sec:3} we describe the data
reduction, and the photometric calibrations. The aperture photometry
is presented in Sect.\ref{sec:4}, whereas Sect. \ref{sec:5} deals with
the LFs and Sect. \ref{sec:6} with the spatial distribution of
galaxies of different luminosity. Sect. \ref{sec:7} is dedicated to
the summary and the discussion of the results. In this work we assume
H$_0$ = 70 \ks Mpc, $\Omega_m$ = 0.3, $\Omega_{\Lambda}$ =
0.7. According to this cosmology, 1 arcmin corresponds to 0.205 Mpc at
z = 0.209.

\section{Observations}
\label{sec:2}

New observations of the galaxy cluster ABCG\,209 were carried out at
the ESO New Technology Telescope (NTT) with the EMMI instrument in
October 2001. The data include B--, V-- and R--band imaging, plus
multi--slit spectroscopy (EMMI--NTT) for 112 cluster members. The
spectroscopic data are presented in Paper I.

   	\begin{figure*}
	 \centering
   	\includegraphics[width=0.7\textwidth]{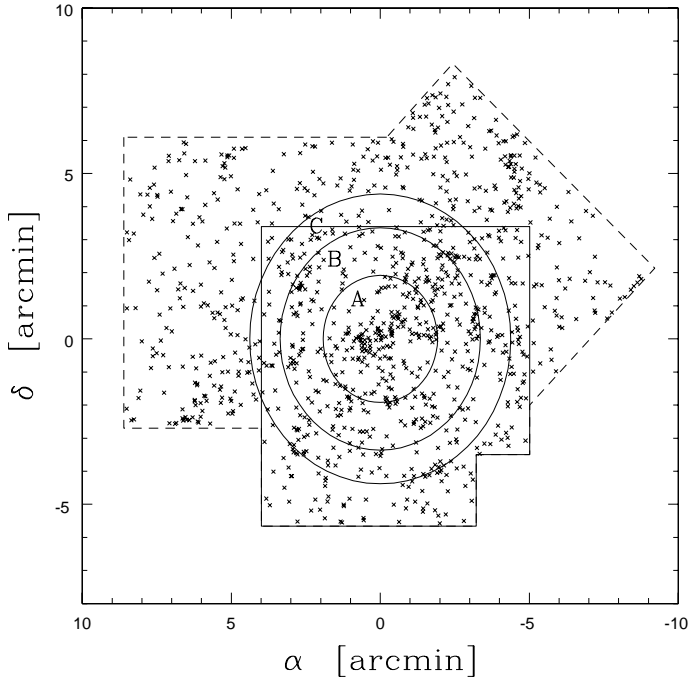}
	\caption{Area covered by the photometry in the region of
   ABCG\,209. All the galaxies brighter than V = 21.5 are marked with
   crosses. The central field (solid contour) was observed in B, V and
   R bands, while the two adjacent fields (dashed contours) were
   observed only in V band. Circles mark the regions analysed in
   Sect. \ref{sec:61}. North is up and East is on the left. The
   origin of the coordinates coincides with the cluster center
   ($\alpha_{2000}$ = 01 31 52.7, $\delta_{2000}$ = -13 36 41.9). A
   region around a bright star was not considered in the analysis
   (right bottom corner of central field).}
	\label{cluster}
	\end{figure*}

A field of $9.2^\prime \times 8.6^\prime$ (1.9 $\times$ 1.8
$\mathrm{h^{-2}_{70}}$ Mpc$^2$), was observed in the B, V and R bands around
the cluster center. In order to sample the cluster at large distance
from the center, we observed other two fields in V band. The total
observed area in V band, accounting for overlapping, is $\sim160$
arcmin$^2$ and is shown in Fig. \ref{cluster}. The relevant
information on the photometry are summarized in Table
\ref{inf}. Standard stars from Landolt (\cite{lan92}) were also
observed before and after the scientific exposures, and were used for
the photometric calibration.

   \begin{table}[h]
      \caption[]{Information on the photometric observations.}
         \label{inf}
     $$
           \begin{array}{c c c c}
            \hline
            \noalign{\smallskip}
            \mathrm{Band} & \mathrm{T_{exp}} & \mathrm{Seeing} & \mathrm{Scale} \\
                 & \mathrm{ks} & \mathrm{arcsec} &\mathrm{arcsec/pxl}\\
            \noalign{\smallskip}
            \hline
            \noalign{\smallskip}
            \mathrm{B} &1.8 & 0.80  & 0.267  \\
            \mathrm{V} &1.26 \times 3 & 0.80 & 0.267   \\
            \mathrm{R} &0.9& 0.90 & 0.267\\
            \noalign{\smallskip}
            \hline
         \end{array}
     $$
   \end{table}

\section{Data reduction and photometric calibration}
\label{sec:3}

Standard procedures were employed for bias subtraction, flat--field
correction, and cosmic ray rejection using the IRAF~\footnote{IRAF is
distributed by the National Optical Astronomy Observatories, which are
operated by the Association of Universities for Research in Astronomy,
Inc., under cooperative agreement with the National Science
Foundation.} package. For each waveband the flat--field was obtained
by combining twilight sky exposures. After bias subtraction and
flat--field correction the images were combined using the IRAF task
IMCOMBINE with the CRREJECT algorithm. Residual cosmic rays and hot
pixels were interpolated applying the IRAF task COSMICRAYS. The
resulting images show a uniform background with typical r.m.s. of
2.3\%, 1.4\%, and 1.1\% for the B, V, and R bands, respectively.

   \begin{table}
      \caption[]{Results of the photometric calibration of BVR
      data.}
         \label{PC}
    $$
   \begin{array}{c c c c c c}
   \hline
   \noalign{\smallskip}
   \mathrm{Band} & \mathrm{C} & \gamma & \mathrm{A} & \mathrm{ZP} & \sigma  \\
   \noalign{\smallskip}
   \hline
   \noalign{\smallskip}
   \mathrm{B} & \mathrm{B - V} & -0.037 \pm 0.008 &  0.322 \pm 0.010 & 24.885 \pm 0.015 & 0.016\\
   \mathrm{V} & \mathrm{B - V} &  0.035 \pm 0.009 &  0.204 \pm 0.013 & 25.277 \pm 0.021 & 0.020\\
   \mathrm{R} & \mathrm{V - R} &  0.010 \pm 0.017 &  0.145 \pm 0.014 & 25.346 \pm 0.023 & 0.026\\
            \noalign{\smallskip}
            \hline
         \end{array}
     $$
   \end{table}

The photometric calibration was performed into the
Johnson--Kron--Cousins photometric system by using the Landolt standard
fields. The instrumental magnitudes of the stars were measured in a
fixed aperture by using the IRAF packages APPHOT and DAOPHOT. The
aperture size was chosen in order to i) enclose the total flux, ii)
obtain the maximum signal--to--noise ratio. By comparing the
magnitudes of the stars in different apertures, we found that a
reasonable compromise is achieved with an aperture of radius 10 pixels
(cf. Howell \cite{how89}). For each band, we adopted the following
calibration relation:

\begin{equation}
\mathrm{M} = \mathrm{M^{'}} + \gamma \cdot \mathrm{C} - \mathrm{A}
\cdot \mathrm{X} + \mathrm{ZP}  \ , 
\label{eq1}
\end{equation}

\noindent
where M and C are the magnitudes and colours of the standard stars,
M$^{'}$ is the instrumental magnitude, $\gamma$ is the coefficient of
the colour term, A is the extinction coefficient, X is the airmass and
ZP is the zero--point. The quantities $\gamma$, A and ZP were derived
by a least square procedure with the IRAF task FITPARAMS. The results
of the photometric calibrations are reported in Table \ref{PC}.
Unless otherwise stated, errors on estimated quantities are given at
68\% confidence level (hereafter c.l.).

As a test we compare the (B--R, V--R) diagram of the Landolt
standard stars with that of the stars in the cluster field. We
measured the magnitude of these stars by using the software SExtractor
(Bertin \& Arnouts \cite{ber96}). We verified that the observed
distribution of stars in our images matches that of the Landolt stars
in the (B--R, B--V) plane, proving the accuracy of the photometric
calibration.

\section{Aperture photometry}
\label{sec:4}

For each image, a photometric catalog was derived by using the
software SExtractor (Bertin \& Arnouts \cite{ber96}). We measured
magnitudes within a fixed aperture of 5.0$^{''}$, corresponding to
$\sim$ 17 kpc at z = 0.209, and Kron magnitudes (Kron \cite{kro80}),
for which we used an adaptive aperture with diameter $a \cdot
r_K$, where $r_K$ is the Kron radius and $a$ is a
constant. We chose $a$ = 2.5, yielding $\sim$ 94\% of the total source
flux within the adaptive aperture (Bertin \& Arnouts
\cite{ber96}). The measured magnitudes were corrected for galactic
extinction following Schlegel et al. (\cite{sch98}). The uncertainties
on the magnitudes were obtained by adding in quadrature both the
uncertainties estimated by SExtractor and the uncertainties on the
photometric calibrations.

   \begin{figure*}[h]
   \centering
   \includegraphics[width=0.7\textwidth]{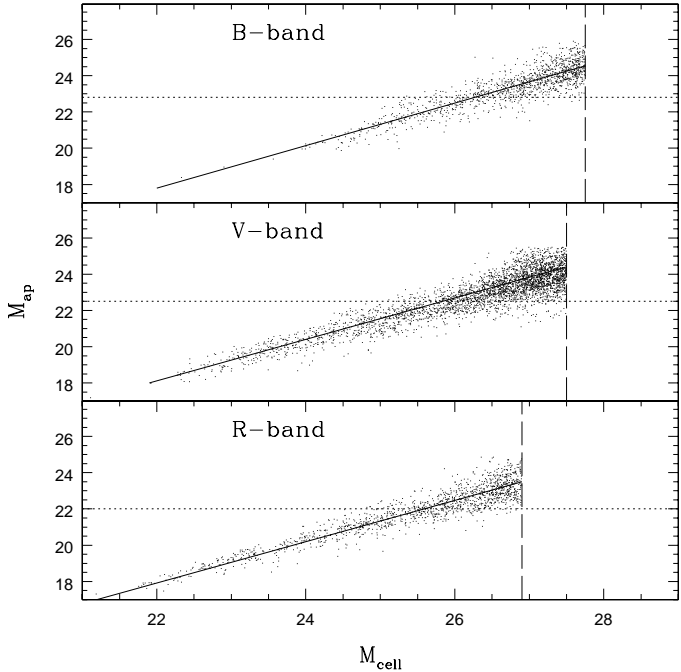}
    \caption{Completeness magnitudes of the B--, V--, and R--band
   images are estimated by comparing magnitudes in the fixed aperture
   (M$_{\mathrm{ap}}$) and in the detection cell
   (M$_{\mathrm{cell}}$). The dotted lines represent the completeness
   limits, the dashed lines mark the limits in the detection cell, and
   solid lines are the linear relations between M$_{\mathrm{ap}}$ 
   and M$_{\mathrm{cell}}$.}
      \label{comple}
    \end{figure*}

The completeness magnitudes were derived following the method of
Garilli et al. (\cite{gar99}), as shown in Fig. \ref{comple}. We
estimated the completeness magnitude as the magnitude where galaxies
start to be lost because they are fainter than the brightness
threshold in the detection cell. The completeness magnitudes are
B$_\mathrm{C}$ = 22.8 (N$\mathrm{_{gal}}$ = 339), V$_\mathrm{C}$ =
22.5 (N$\mathrm{_{gal}}$ = 1078), and R$_\mathrm{C}$ = 22.0
(N$\mathrm{_{gal}}$ = 679).

In order to derive the LF we considered the objects brighter than the
completeness limit and adopted the Kron magnitude, since this is the
best estimate of the total magnitude. The star/galaxy classification
was based on the SExtractor stellar index (SG), defining as stars the
sources with SG $\geq$ 0.98.

The catalogs for the central region of the cluster were
cross--correlated by using the IRAF task XYXYMATCH. For each object,
colours were derived within the fixed aperture.

\section{Luminosity Functions}
\label{sec:5}

In order to measure the cluster LF in each band we used all the galaxy
photometric data up to the completeness magnitude, and removed the
interlopers by statistically subtracting the background
contamination. We used galaxy counts in B, V and R bands from the
ESO--Sculptor Survey (Arnouts et al. \cite{arn97}; de Lapparent et
al. \cite{del03}) kindly provided to us by V. de Lapparent. These data
cover an area of $\sim$ 729 arcmin$^2$ observed with EMMI
instrument. The data reduction was performed by using the procedures
similar to those adopted in the present work. In particular, the
photometric catalog was obtained by SExtractor and the total
magnitudes were estimated from the Kron magnitudes defined by adopting
the same aperture.

We assumed Poisson statistics for the background and cluster field
galaxy counts. The errors on the cluster LFs were computed by adding
in quadrature Poissonian fluctuations. Since the field of view of our
observations is $\sim$ 10 times smaller than the area covered by the
background counts, the errors on the cluster LFs are dominated by
Poissonian errors of cluster counts.

We derived the LFs for the central field (see Fig. \ref{cluster}) in
the B, V and R bands by fitting the galaxies counts with a single
Schechter function. In V and R bands the fits were also computed with
the sum of two Schechter functions in order to describe the bright and
the faint populations (Sect. \ref{sec:51}). We compared the LFs with
counts obtained selecting red sequence galaxies
(Sect. \ref{sec:52}). All the fit parameters and the $\chi^2$
statistics, are listed in Table \ref{fitsLF}.

\subsection{Multiband analysis}
\label{sec:51}

\begin{figure*} 
   \vspace{-2cm}
   \centering
   \includegraphics[width=1.0\textwidth]{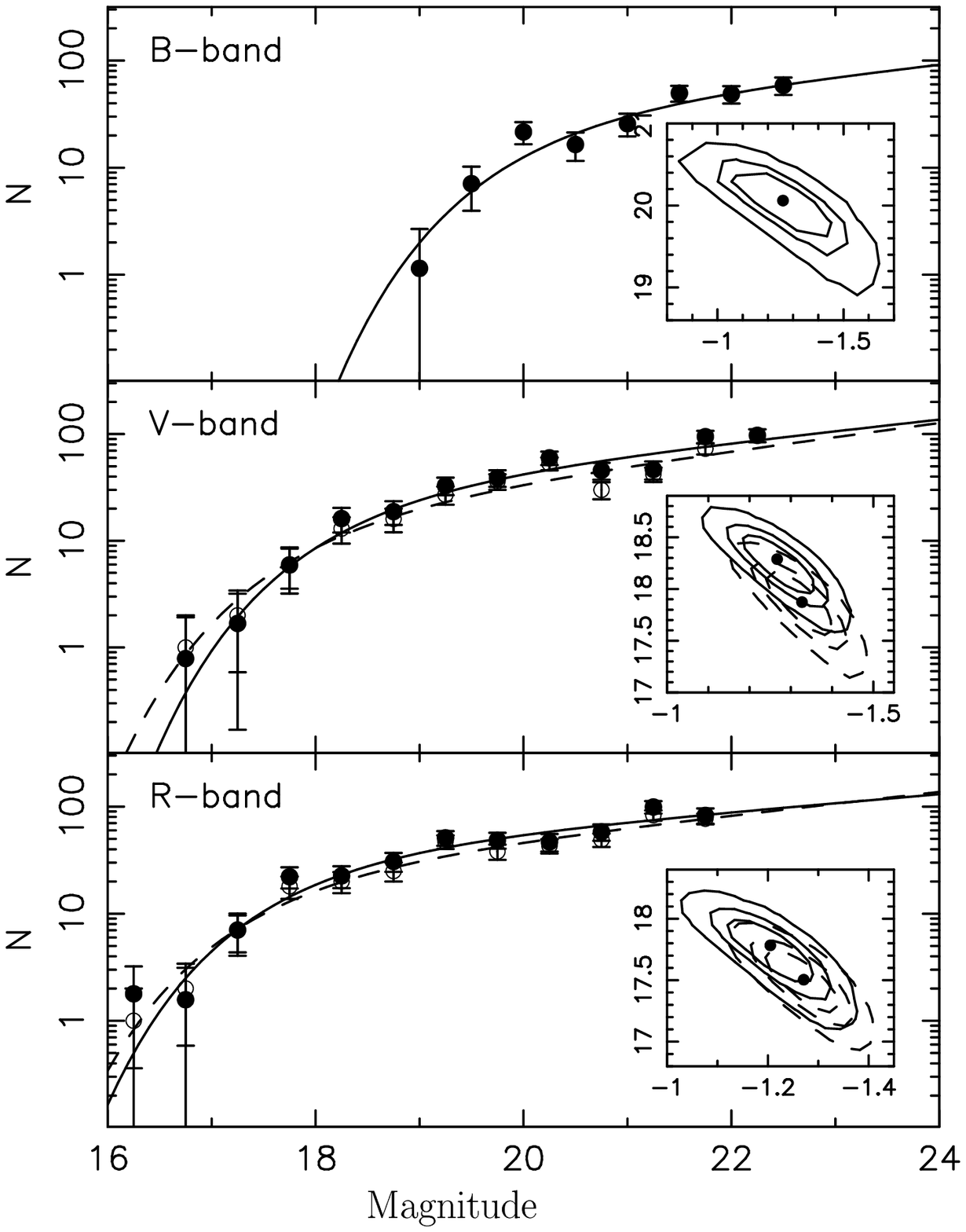}
   \vspace{-4cm} 
   \caption{Luminosity function in the B, V and R bands in the central
   field of 9.2$^\prime \times$ 8.6$^\prime$. In V and R bands filled
   circles represent counts obtained from the photometric catalog with
   a statistical background subtraction, open circles are the counts
   of galaxies belonging to the red sequence of the CM relation (see
   Sect. \ref{sec:52}). Solid and dashed lines are the Schechter
   fits to the filled and open circles respectively. In the small
   panels the 1, 2 and 3$\sigma$ c.l.  of the best--fit parameters
   $\alpha$ and M$^*$ are shown.}
   \label{centLF} 
   \end{figure*}

Figure \ref{centLF} (solid lines) shows the LFs in the B, V and R
bands for the central cluster region, obtained by a weighted
parametric fit of the Schechter function to the statistically
background--subtracted galaxy counts (filled circles). The parameters
of the fit are: B$^*$ = 20.06, $\mathrm{\alpha_B}$ = -1.26, V$^*$ =
18.29, $\mathrm{\alpha_V}$ = -1.27, R$^*$ = 17.78, $\mathrm{\alpha_R}$
= -1.20. We evaluated the quality of the fits by means of the $\chi^2$
statistics (see Table \ref{fitsLF}). The single Schechter function
gives a fair representation of the global distribution of the data,
that is the single Schechter fit cannot be rejected in any band even
at the $10\%$ c.l. .

   \begin{table} 
     \caption[]{Fits to the Luminosity Functions. Errors on the
      $\mathrm{M^*}$ and $\alpha$ parameters can be obtained by the
      confidence contours shown in Fig. \ref{centLF}.}
\label{fitsLF}
     $$
           \begin{array}{c| c c c c| c c c c }
            \hline
            \noalign{\smallskip}
            \mathrm{Band} & \multicolumn{4}{c|}{\mathrm{~~~~~Single~Schechter~function}~~~~~} & \multicolumn{4}{c}{\mathrm{~~~~~Two~Schechter~functions^{(a)}~}}\\
		\noalign{\smallskip}
		\hline
		\noalign{\smallskip}

 		 &~~~~~~\mathrm{M^*}~~~~~~&~~~~\alpha~~~~&~~~~~\mathrm{\chi^{2(b)}_{\nu}}~~~~~&\mathrm{P(\chi^2>\chi^2_{\nu})}~&~~~\mathrm{M}^*_{faint}~~~&~~\alpha_{faint}~~&~~~\mathrm{\chi^2_{\nu}}~~~&\mathrm{P(\chi^2>\chi^2_{\nu})}\\

            \noalign{\smallskip}
            \hline
            \noalign{\smallskip}
            \mathrm{B} & -21.03 & -1.26 & 0.96 & 46\% &        &       &      &      \\
            \mathrm{V^{c}} & -22.03 & -1.25 & 1.04 & 40\% &        &       &      &      \\
            \mathrm{V} & -22.18 & -1.27 & 1.46 & 15\% & -18.72 & -2.00 & 1.09 & 37\% \\
            \mathrm{R} & -22.48 & -1.20 & 1.24 & 27\% & -19.14 & -1.24 & 1.06 & 39\% \\
            \noalign{\smallskip}
            \hline
            \noalign{\smallskip}
            \multicolumn{1}{c} {\mathrm{    }} & \multicolumn{4}{c}{\mathrm{~~~~~Galaxies~on~the~red~sequence}~~~~~} & \multicolumn{4}{c}{}\\
            \noalign{\smallskip}
            \hline
            \mathrm{V} & -22.60 & -1.33 & 2.10 &  2\% &   -18.68   &  -2.19  & 1.68      &  9 \%   \\
            \mathrm{R} & -22.76 & -1.27 & 1.30 & 23\% &   -19.15     &  -1.39     &  1.09    &  37 \% \\
            \noalign{\smallskip}
            \hline
         \end{array}
    $$
\begin{list}{}{}  
\item[$\mathrm{^{a}}$] The bright--end LF is fixed (see text).
\item[$\mathrm{^{b}}$] The reduced $\mathrm{\chi^2}$.
\item[$\mathrm{^{c}}$] Total observed field (see Sect. \ref{sec:6}).
\end{list}

   \end{table}

On the other hand, there is indication of a dip in the
distribution at V $\sim$ 21.0 and R $\sim$ 20.5.  According to the
fitted single Schechter function, there should be 121 and 125 galaxies
in the range V = 20.5--21.5 and R = 20.0--21.0 respectively, whereas
in our counts we find $\sim 92 \pm 10$ and $\sim 105 \pm 10$
galaxies. If we define the dip amplitude as:

\begin{equation}
\mathrm{
A=\frac{N_e-N_o}{N_e} \ ,
}
\label{dipa}
\end{equation}

\noindent
where $\mathrm{N_e}$ and $\mathrm{N_o}$ are the expected and observed
number of galaxies in the dip magnitude range, we obtain in V (R) band
A$=24 \pm 8 \%$ ($16 \pm 8 \%$).  The position of dips are: M$\mathrm{_V}$
$\sim$ -19.5 and M$\mathrm{_R}$ $\sim$ -19.8. These values were obtained by
converting apparent into absolute magnitudes using k--corrections for
early--type galaxies from Poggianti (\cite{pog97}). The B--band data
are not deep enough to sample the dip position.
 
The presence of a dip was discussed in the literature by considering
the properties of the TSLFs (e.g., Binggeli et al. \cite{bin88}) in
nearby clusters. In a study of the Virgo cluster, Sandage et
al. (\cite{san85}) showed the presence of two distinct classes of
galaxies (normal and dwarfs), with different dynamical and luminosity
evolution. Since the bright--end of the LF is well studied, we can use
{\it a priori} the information about the LF shape for bright galaxies
to fit our data with two different Schechter functions, representing
bright and faint galaxies. We assumed a Schechter model with
R$^*\mathrm{_{bright}}$ = 18.0 and $\alpha\mathrm{_{R,bright}}$ = -1.0
(from N$\mathrm{\ddot{a}}$slund et al. \cite{nas00} results, scaled
according to the adopted cosmology). Using the V--R colour term as
derived from Eq.~(\ref{EQCM}), we also fix V$^*\mathrm{_{bright}}$ =
18.7 and $\alpha\mathrm{_{V,bright}}$ = -1.0.

\begin{figure*} 
   \vspace{-2cm}
   \centering
   \includegraphics[width=1.0\textwidth]{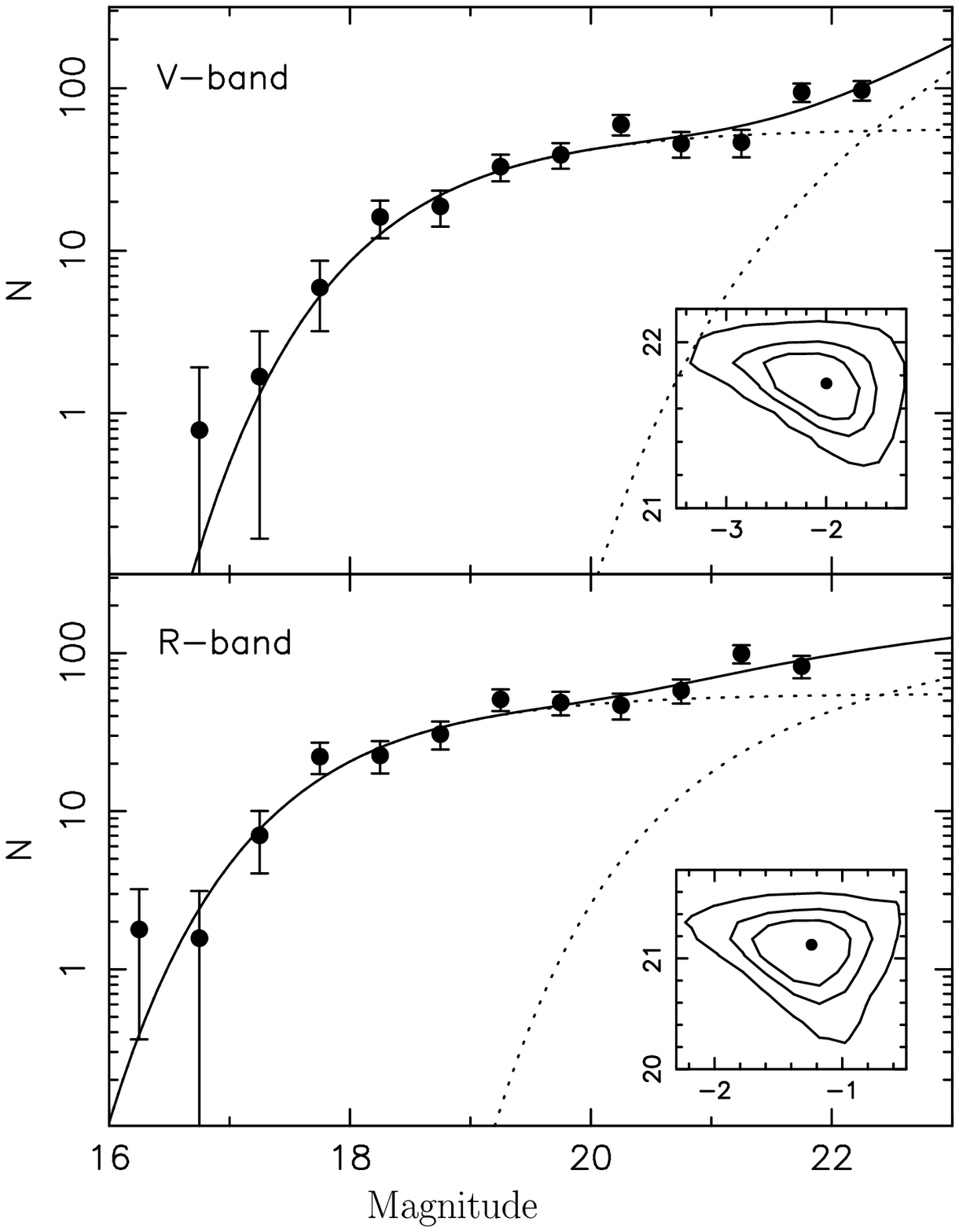}
   \vspace{-4.0cm}
   \caption{Luminosity function in the V and R bands in the central
            field of 9.2$^\prime \times$ 8.6$^\prime$. The dashed
            lines represent the Schechter functions for bright and
            faint galaxies separately (see text). The continuous line
            represents the sum of the two functions. In the small
            panels the 1, 2 and 3$\sigma$ c.l. of the best--fit
            parameters $\alpha$ and M$^*$ are shown.}
            \label{2LF} \end{figure*}

Figure \ref{2LF} shows the cluster LFs in V and R bands modelled with
the two Schechter functions. The fit procedures yields
V$^*\mathrm{_{faint}}$ = 21.75, R$^*\mathrm{_{faint}}$ = 21.12. For
the slope of the faint galaxies function the V-- and R--band data
provide poor constraints, so that we have only an indication of a
steep faint--end. According to the $\chi^2$ statistics (see Table
\ref{fitsLF}), combining two Schechter functions for bright and faint
galaxies the quality of the fit increases both in V and R bands.

\subsection{Galaxies on the red sequence}
\label{sec:52}

\begin{figure*}
\centering
\includegraphics[width=0.8\textwidth]{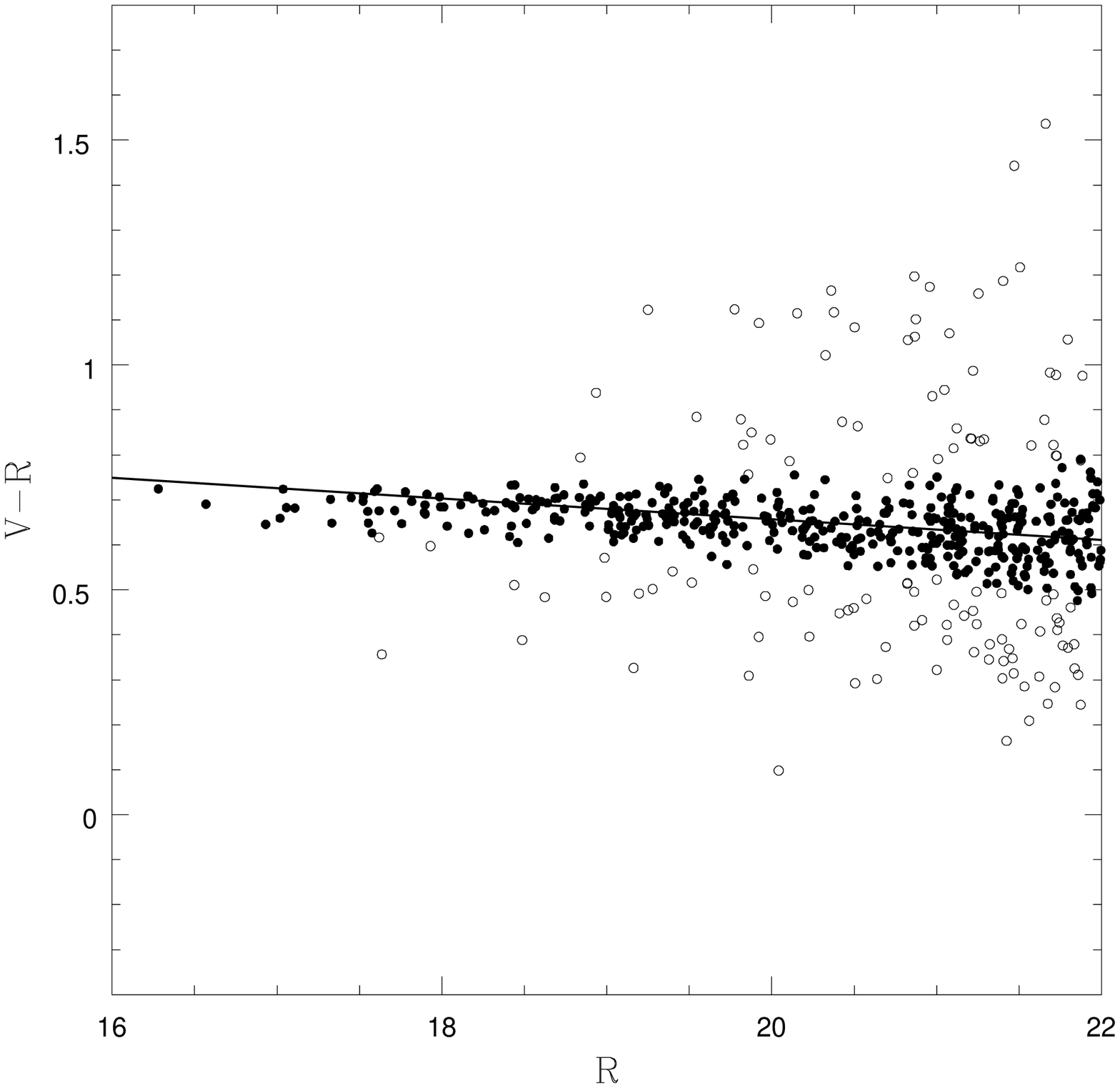}
\caption{V--R vs. R CM diagram for all the galaxies within the
completeness limit in the central field of ABCG\,209. Galaxies of the
red sequence (see solid line) are plotted as filled circles. The
solid line defines the CM sequence for spectroscopically confirmed
cluster members.}
\label{CM}
\end{figure*}

We obtained the Colour--Magnitude (CM) relation by fitting the
photometric data of the spectroscopically confirmed cluster members
(see Paper I) with a biweight algorithm (Beers et al. \cite{bee90}):

\begin{equation}
\mathrm{
(V-R)_{CM} = - 0.023 \cdot R + 1.117 \ .
}
\label{EQCM}
\end{equation}
\noindent
By using Eq.~(\ref{EQCM}), we defined as sequence galaxies the sources
lying in the region inside the curves:

\begin{equation}
\mathrm{
(V-R)_{seq} = (V-R)_{CM} \pm (\sqrt{\sigma_V^2 +\sigma_R^2} + 0.05) \ ,
}
\label{EQCMS}
\end{equation}
\noindent
where we took into account the photometric uncertainty at 1$\sigma$
both on the V ($\mathrm{\sigma_V}$) and on the R magnitude
($\mathrm{\sigma_R}$) as well as the intrinsic dispersion of the CM relation
(Moretti et al. \cite{mor99}). In Fig. \ref{CM} the red sequence
galaxies are marked with filled circles.

By selecting galaxies on the CMR (within the observed scatter), the
background contamination is expected to be negligible.  So we can
directly compare galaxy counts (open circles in Fig. \ref{centLF})
with those derived in Sect. \ref{sec:51} (filled circles in
Fig. \ref{centLF}). In both V and R bands the counts obtained with the
two different approaches are very similar.

Figure \ref{centLF} (dashed lines) shows the LFs in the V and R
bands for the central cluster region, obtained by a weighted
parametric fit of the Schechter function to the red galaxy counts
(open circles). The parameters of the fit are: V$^*$ = 17.87,
$\mathrm{\alpha_V}$ = -1.33, R$^*$ = 17.50, $\mathrm{\alpha_R}$ =
-1.27. According to the $\chi^2$ statistics (see Table \ref{fitsLF})
we can reject the fit with a single Schechter function in the V band
at 98\% c.l. At the same time, the dip amplitude in the V band, A$=26
\pm 9 \%$, is unchanged respect to the case of the global LF
(Sect. \ref{sec:51}). It turns out that, independently from the fitted
Schechter functions, the ratio ($\sim 60 \%$) of observed counts in
the bins inside the dip and in the bins adjacent to the dip regions is
the same for both galaxy samples, while the overall distribution of
red galaxies cannot be described with a single Schechter
function. According to the $\chi^2$ statistics (see Table
\ref{fitsLF}), combining two Schechter functions the quality of the
fit increases both in V and R bands, and becomes acceptable also in
the V band.

\section{Luminosity segregation}
\label{sec:6}

The data in the V band, covering an area of $\sim$ 160 arcmin$^2$,
corresponding to a circular region with equivalent radius 0.6
R$_{\mathrm{vir}}$ (R$_{\mathrm{vir}}$ = 2.5 h$^{-1}_{70}$ Mpc; see
Paper I), allow to study the environmental dependence of LF and the
spatial distribution of galaxies as a function of the clustercentric
distance.

Figure \ref{LFsegr_fit} (upper panel) shows the V--band LF in the
whole observed area, modelled by using a weighted parametric fit to a
single Schechter function, with best fit values V$^*$ = 18.45
(M$\mathrm{^*_V}$ = -22.03) and $\mathrm{\alpha_V}$ = -1.25. The LF shape is
very similar to that obtained in the central field (Fig. \ref{centLF})
and also in this case the Schechter function overestimates the
observed counts in the range V = 20.5--21.5. The dip amplitude is
A$=14 \pm 7 \%$.

\subsection{The LF in different cluster regions}
\label{sec:61}

   \begin{table}
     \caption[]{Best fit value of the LFs measured in three regions
      around the center of the cluster.} 
    \label{LFsegr}
     $$
           \begin{array}{c c c c c c }
            \hline
            \noalign{\smallskip}
    \mathrm{Region} & \mathrm{Area (h^{-2}_{70} Mpc^2)} & \mathrm{V}^* & \mathrm{Luminosity \ ratio}& \mathrm{\chi^2_{\nu}} &\mathrm{P(\chi^2>\chi^2_{\nu})} \\
            \noalign{\smallskip}
            \hline
            \noalign{\smallskip}
    \mathrm{A} & 0.50   & 18.09 \pm 0.73 & 12.7 \pm 3.0 & 1.12 & 35\%\\
    \mathrm{B} & 1.00   & 18.49 \pm 0.44 & 6.9  \pm 1.4 & 1.08 & 37\% \\
    \mathrm{C} & 1.00   & 18.88 \pm 0.43 & 6.8  \pm 2.3 & 0.85 & 56\% \\
            \noalign{\smallskip}
            \hline
         \end{array}
     $$

   \end{table}

\begin{figure*}
   \vspace{-2cm} \centering
   \includegraphics[width=1.0\textwidth]{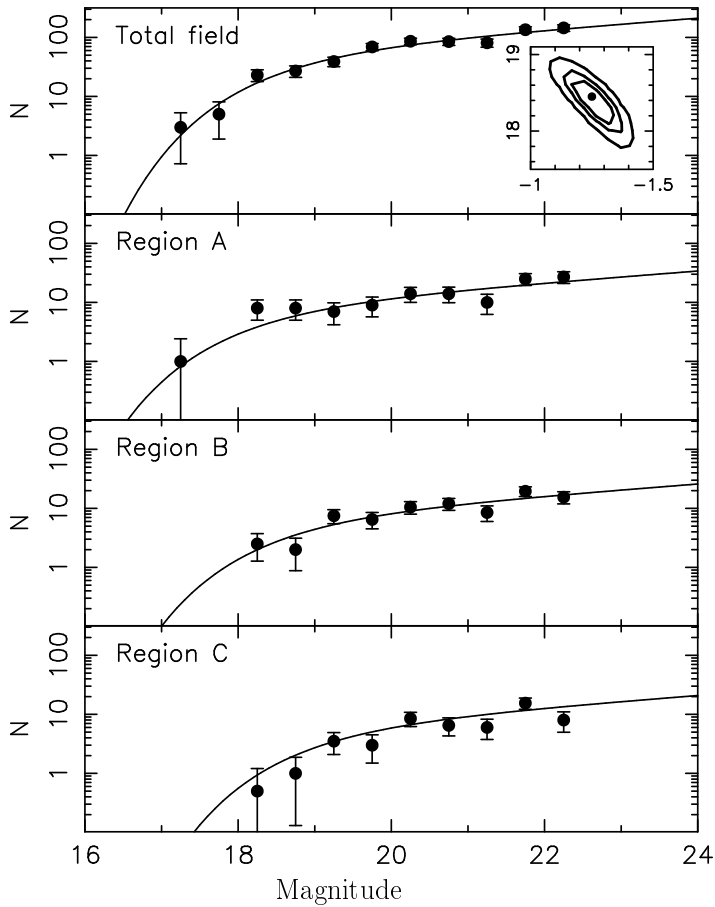}
   \vspace{-4.0cm}
    \caption{Luminosity function in the V band. In the upper panel are
    shown the LF in the whole observed area (160 arcmin$^2$) and the
    1, 2 and 3$\sigma$ c.l. of the best--fit parameters $\alpha$ and
    M$^*$ (small panel). The continuous line represents the fitted
    Schechter functions. In the other panels the LFs in the three
    cluster regions (see text) are shown. Continuous lines are the fit
    of the Schechter functions obtained by fixing the faint--end slope
    at the best fit value of the LF fitted on the whole observed
    area.}
     \label{LFsegr_fit}
    \end{figure*}

In order to compare the abundances of galaxies of various luminosities
in different regions we computed the LFs in three areas at different
distances from the center~\footnote{The center of the cluster was
derived by a two--dimensional adaptive kernel technique from the
spectroscopically confirmed cluster members (for details see PaperI)
and coincides with the cD position.}. First we considered the cluster
counts in an area of 0.5 h$^{-2}_{70}$ Mpc$^2$ around the center of
the cluster (region A), and then in two concentric circular rings
around the first central region, each in an area of 1.0 h$^{-2}_{70}$
Mpc$^2$, respectively at $\sim$ 2$^\prime$ and $\sim$ 4.7$^\prime$
from the center (region B and C). We fitted the counts using a single
Schechter function with $\alpha$ fixed at the best fit value obtained
from the LF computed over the whole observed field. Table \ref{LFsegr}
reports the relative fit values in the different regions and
Fig. \ref{LFsegr_fit} shows the fitted functions. We also measured the
luminosity weighted ratio of the number of objects brighter and
fainter than the V = 21.0 (Table \ref{LFsegr}), that is the magnitude
where the dip occurs. This luminosity ratio increases from region
B to the center by a factor $1.8 \pm 0.6$, indicating a significant
luminosity segregation, whereas the ratio does not vary from the
region B to the region C.

\subsection{Spatial distribution of galaxies}
\label{sec:62}

\begin{figure*}
   \includegraphics[width=0.5\textwidth,angle=-90]{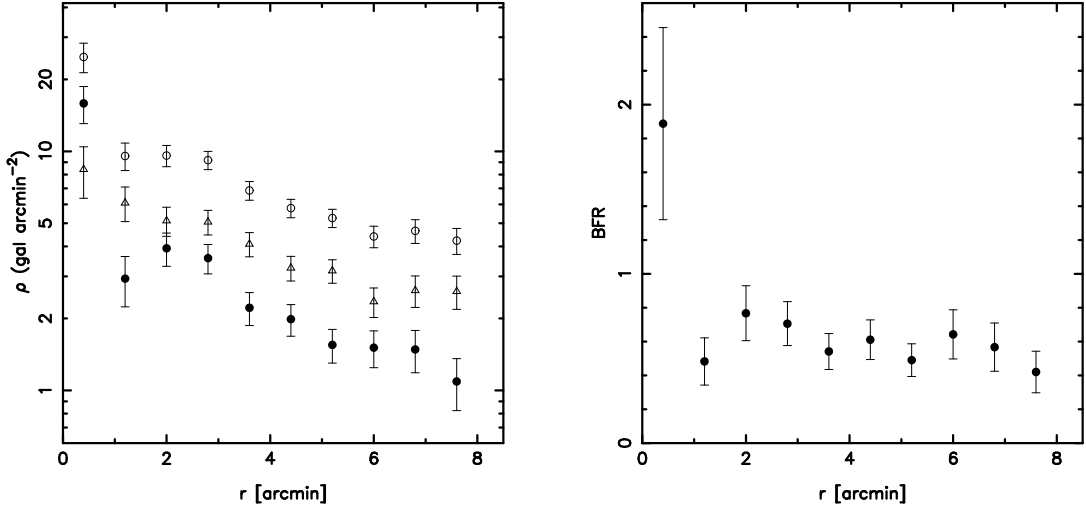}
    \caption{The background--subtracted radial profile of ABCG\,209 is
    shown in the left panel. Open circles represent the distribution
    of all galaxies in the cluster, filled circles are bright galaxies
    (M$\mathrm{^*_V \le }$ -19.5) and open triangles are faint
    galaxies (M$\mathrm{^*_V > }$ -19.5). The number of bright
    galaxies in the first and in the second bin is $\sim$ 32 and
    $\sim$ 17 respectively, while the number of faint galaxies is
    $\sim$ 15 and $\sim$ 30. The bright--to--faint ratio as a
    function of the clustercentric distance is shown in the right
    panel.}
      \label{pdens}
    \end{figure*}

In Fig. \ref{pdens} (left panel) the number density radial profile is
shown, i.e. the number of galaxies measured in concentric rings around
the center. It has been computed for the whole galaxy population (open
circles) and for galaxies brighter (filled circles) and fainter
(triangles) than V = 21.0, the magnitude where the luminosity
distribution shows the distinctive upturn. The number of faint
galaxies is determined up to the completeness limit. The densities
are statistically background--subtracted by using background galaxy
density measured in the ESO--Sculptor Survey (Arnouts et
al. \cite{arn97}). Errors are assumed to be Poissonian.

The radial profile of all galaxies is centrally peaked, showing the
highest gradient for bright galaxies (Fig. \ref{pdens}, left
panel). Indeed, the bright galaxy density decreases by a factor $\sim
5.4$ between the first and the second bin, to be compared with a
decrease of a factor $\sim 1.4$ for faint galaxies and of a factor
$\sim 2.6$ when all galaxies are considered. For faint galaxies (V $>$
21.0), the distribution decreases slowly with clustercentric distance.

Fig. \ref{pdens} (right panel) shows the bright--to--faint ratio (BFR)
as a function of the clustercentric distance. The BFR shows a maximum
at the center, where the number of bright galaxies is two times the
number of faint galaxies, and decreases rapidly between the first and
the second bin. Outside the central region of the cluster (r $>$ 1
arcmin) the BFR distribution is flat and there are roughly two faint
galaxies for each bright galaxy.

\section{Summary and discussion}
\label{sec:7}

The analysis performed on ABCG\,209 can be summarized by the following
results:

\begin{description}

\item[1)] the fit with a single Schechter function cannot be
rejected in any band; at the same time, the luminosity distributions
are better described by a sum of two Schechter functions for bright
and for faint galaxies respectively;

\item[2)] there is an indication of a dip in V-- and R--band LFs,
and this features is also present when considering only galaxies that
lie on the red sequence ;

\item[3)] the faint--end slope seems to be steeper going from redder
to bluer wavelengths;

\item[4)] there is a luminosity segregation, in the sense that the
V$^*$ value increases from the center to the outer regions and the
luminosity ratio of bright--to--faint galaxies decreases by a factor
$\sim 2$ from the center to outer regions, whereas the number ratio
decreases by a factor $\sim 4$.

\end{description}

The presence of a dip was found in the LF of many clusters. In Table
\ref{DIP} we reported the positions of the dips for some of these
clusters with 0.0$<$ z $<$ 0.3 in R--band absolute magnitudes
(according to the adopted cosmology). The literature data were
transformed to R--band absolute magnitudes by using k-- and
evolutionary correction, and average colours for early--type galaxies
of Poggianti (\cite{pog97}).

   \begin{table}

      \caption[]{Compilation of dip positions in the restframe R band
      in cluster LFs from literature. Data were transformed to R--band
      absolute magnitudes by using k-- and evolutionary correction,
      and average colours for early--type galaxies of Poggianti
      (\cite{pog97}).}

         \label{DIP}
     $$
           \begin{array}{c c c  c }
            \hline
            \noalign{\smallskip}
             \mathrm{Cluster \ name} & \mathrm{Redshift}  &  \mathrm{~~~~Dip \ position} & \mathrm{Reference} \\
            \noalign{\smallskip}
            \hline
            \noalign{\smallskip}
            \mathrm{Virgo}        & 0.0040 & -18.9 & \mathrm{Sandage \ et \ al. (\cite{san85})} \\
            \mathrm{ABCG\,194}    & 0.018  & -19.8 & \mathrm{Yagi \ et \ al. (\cite{yag02})}\\
            \mathrm{Coma}         & 0.0232 & -19.8 & \mathrm{Biviano \ et \ al. (\cite{biv95})} \\
            \mathrm{ABCG\,496}    & 0.0331 & -19.8 & \mathrm{Durret \ et \ al. (\cite{dur00})} \\
            \mathrm{ABCG\,2063}   & 0.035  & -19.3 & \mathrm{Yagi \ et \ al. (\cite{yag02})} \\
            \mathrm{ABCG\,576}    & 0.038  & -18.6 & \mathrm{Mohr \ et \ al. (\cite{moh96})} \\
            \mathrm{Shapley 8}    & 0.0482 & -19.5 & \mathrm{Metcalfe \ et \ al. (\cite{met94})} \\
            \mathrm{ABCG\,754}    & 0.053  & -19.3 & \mathrm{Yagi \ et \ al. (\cite{yag02})} \\
            \mathrm{ABCG\,85(z)}^{\mathrm{a}}   & 0.0555 & -19.9 & \mathrm{Durret \ et \ al. (\cite{dur99})} \\
            \mathrm{ABCG\,85} & 0.0555 & -19.3 & \mathrm{Durret \ et \ al. (\cite{dur99})} \\
            \mathrm{ABCG\,2670}   & 0.076  & -19.3 & \mathrm{Yagi \ et \ al. (\cite{yag02})} \\
            \mathrm{ABCG\,1689}   & 0.181  & -18.7 & \mathrm{Wilson \ et \ al. (\cite{wil97})} \\
            \mathrm{ABCG\,665}& 0.182  & -18.7 & \mathrm{Wilson \ et \ al. (\cite{wil97})} \\
            \mathrm{ABCG\,963}& 0.206  & -19.0 & \mathrm{Driver \ et \ al. (\cite{dri94})} \\
            \mathrm{ABCG\,209}& 0.209  & -19.6 & \mathrm{This \, paper}           \\
            \mathrm{MS\,2255.7+2039} & 0.288 & -19.6 & \mathrm{N\ddot{a}slund \ et \ al. (\cite{nas00})}           \\

            \noalign{\smallskip}
            \hline
         \end{array}
     $$
\begin{list}{}{}  
\item[$\mathrm{^{a}}$] LF fitted only on the spectroscopically
confirmed cluster galaxies.
\end{list}
   \end{table}

Dips are found at comparable absolute magnitudes, in different
clusters, suggesting that these clusters may have similar galaxy
population. However, the dips have not necessarily the same shape:
the dips found in the the LF of Shapley 8 and Coma are broader
than those found in Virgo, ABCG\,963, and ABCG\,85. As already
mentioned in the introduction, this can be explained with the
hypothesis of an universal TSLF (Binggeli et al. \cite{bin88}),
whereas the different relative abundances of galaxy types, induced by
cluster--related processes, are at the origin of the different dip
shapes (see also Andreon \cite{and98}).

Moreover, as shown in Table \ref{DIP}, there is no correlation between
the dip position and the redshift, suggesting little or no
evolutionary effect in the dip, in agreement with
N$\mathrm{\ddot{a}}$slund et al. (\cite{nas00}), which directly
compare the LFs of Coma, ABCG\,963 and MS2255. They detected no
qualitative difference between nearby and distant clusters.

Studying the R--band LFs for a photometric sample of 10 clusters at
different redshifts and with different richness classes, Yagi et
al. (\cite{yag02}) demonstrated that the dips seen in the LFs are
almost entirely due to r$^{1/4}$--like galaxies. This evidence is in
agreement with those shown in Fig. \ref{centLF} (open circles).  In
fact by comparing the counts of red sequence galaxies with those
obtained with a statistical background subtraction, the dip in V band
appears more pronounced.

Environmental effects can be at the origin of the different positions
of the dips seen in different clusters, since the transition from the
bright--dominated to the faint--dominated parts of the LF can occur at
different magnitudes in different environment. The cluster
morphological mix and the morphology--density relation (Dressler
\cite{dre80}) should give rise to LFs with different shape when
subdividing a sample galaxies respect to the cluster environments.

The effect of the environment can be seen in ABCG\,209 by comparing
the shape of the LF in V band in different regions around the cluster
center (see Fig. \ref{LFsegr_fit} and Sect. \ref{sec:61}). The fitted
M$^*$ value is shifted toward fainter magnitudes going from the inner
(A) to the outer (C) region and the luminosity ratio of
bright--to--faint galaxies decreases by a factor $\sim 2$, indicating
a luminosity segregation.

The bright galaxies are markedly segregated in the inner 0.2
h$^{-1}_{70}$, around the cD galaxy. This suggest that bright galaxies
could trace the remnant of the core--halo structure of a pre--merging
clump.

Although ABCG\,209 is a cD--like cluster, with cD galaxy located in
the center of a main X--ray peak, it shows an elongation and asymmetry
in the galaxy distribution (Paper I). Moreover, the faint--end slope
turns out to be $\alpha < -1$ at more than 3$\sigma$ c.l. in both V
and R bands, thus reconciling the asymmetric properties of X--ray
emission with the non flat--LF shape of irregular systems as found by
Lopez--Cruz et al. (\cite{lop97}).

These results allow to discriminate between the two possible formation
scenarios suggested by the dynamical analysis (Paper I). We conclude
that ABCG\,209 is an evolved cluster, resulting from the merger of two
or more sub--clusters, while the elongation and asymmetry of the galaxy
distribution (of the X--ray emission) and the shape of the LFs show
that ABCG\,209 is not yet a fully relaxed system.

Our analysis is in agreement with the existence of i) an universal LF
for bright galaxies, which is well described by a flat Schechter
function, and ii) a steep Schechter function for faint galaxies. A
definitive conclusion regarding the faint--end slope of the LF needs
deeper photometry able to sample the luminosity distribution of dwarf
galaxies.

\begin{acknowledgements}
We thank Valerie de Lapparent who provided us with the galaxy counts
used to derive the cluster LFs. A. M. thanks Massimo Capaccioli for
the hospitality at the Osservatorio Astronomico di Capodimonte, and
Francesca Matteucci for support during this work. This work has been
partially supported by the Italian Ministry of Education, University,
and Research (MIUR) grant COFIN2001028932: {\it Clusters and groups of
galaxies, the interplay of dark and baryonic matter}, and by the
Italian Space Agency (ASI).
\end{acknowledgements}

\end{document}